\documentclass[sigconf]{acmart}

\usepackage[lined,noend,linesnumbered]{algorithm2e}
\usepackage{caption,subcaption}
\usepackage{mathtools}
\usepackage{titlesec}
\titlespacing*{\section}      {0em}{.1em}{.5em}
\titlespacing*{\subsection}   {0em}{-\parskip}{.25em}

\DeclarePairedDelimiter\floor{\lfloor}{\rfloor}

\SetKwInput{KwData}{Inputs}

\SetKwInput{KwResult}{Output}

\copyrightyear{2021}
\acmYear{2021}
\setcopyright{acmlicensed}\acmConference[HPDC '21]{Proceedings of the 30th International Symposium on High-Performance Parallel and Distributed Computing}{June 21--25, 2021}{Virtual Event, Sweden}
\acmBooktitle{Proceedings of the 30th International Symposium on High-Performance Parallel and Distributed Computing (HPDC '21), June 21--25, 2021, Virtual Event, Sweden}
\acmPrice{15.00}
\acmDOI{10.1145/3431379.3460646}
\acmISBN{978-1-4503-8217-5/21/06}



\begin{document}
\fancyhead{}

\title{LaSS: Running Latency Sensitive Serverless Computations at the Edge}

\author{Bin Wang}
\affiliation{%
  \institution{University of Massachusetts Amherst}
  \city{}
  \country{}
}
\email{binwang@cs.umass.edu}

\author{Ahmed Ali-Eldin}
\affiliation{%
  \institution{Chalmers University of Technology}
  \city{}
  \country{}
}
\email{ahmed.hassan@chalmers.se}

\author{Prashant Shenoy}
\affiliation{%
  \institution{University of Massachusetts Amherst}
  \city{}
  \country{}
}
\email{shenoy@cs.umass.edu}

\begin{abstract}
  Serverless computing has emerged as a new paradigm for running short-lived computations in the
  cloud. Due to its ability to handle IoT workloads, there has been considerable interest in running
  serverless functions at the edge. However, the constrained nature of the edge and the latency
  sensitive nature of workloads result in many challenges for serverless platforms. In this paper,
  we present LaSS, a platform that uses model-driven approaches for running latency-sensitive
  serverless computations on edge resources. LaSS uses principled queuing-based methods to determine
  an appropriate allocation for each hosted function and auto-scales the allocated resources in
  response to workload dynamics. LaSS uses a fair-share allocation approach to guarantee a minimum
  of allocated resources to each function in the presence of overload. In addition, it utilizes
  resource reclamation methods based on container deflation and termination to reassign resources
  from over-provisioned functions to under-provisioned ones. We implement a prototype of our
  approach on an OpenWhisk serverless edge cluster and conduct a detailed experimental evaluation.
  Our results show that LaSS can accurately predict the resources needed for serverless functions in
  the presence of highly dynamic workloads, and reprovision container capacity within hundreds of
  milliseconds while maintaining fair share allocation guarantees.
\end{abstract}

\maketitle

\section{Introduction}
\label{sec:intro}

Serverless computing has emerged as a new paradigm of cloud computing for running short-lived
computations. Unlike traditional Infrastructure-as-a-Service (IaaS) clouds \cite{fox2009above} that
involves leasing server resources for longer time periods, serverless computing offers a
Function-as-a-Service (FaaS) abstraction \cite{castro2017serverless} that is designed for infrequent
or bursty short-lived cloud computations. The serverless model involves running code in the form of
functions that are dynamically invoked by the cloud platform for execution upon the arrival of new
requests or data to an application. The invoked function runs often inside a virtualized container
on a cloud server to process a request, after which the container
terminates~\cite{shahrad2020serverless}.

Serverless computing is attractive for application development for two reasons. In case of
infrequent requests, it eliminates the need to lease persistent cloud servers that stay idle between
successive requests. Serverless platforms only charge for the CPU time spent in executing incoming
requests and do \textit{not} charge for idle cycles. Second, in case of bursty computations, it
eliminates the need to pre-provision cloud servers to handle the \emph{peak} load during a burst and
also avoids server overloads if the burst of requests exceeds the provisioned capacity. Serverless
platforms offer the abstraction of an ``infinitely scalable'' cloud through built-in auto-scaling
with the number of containers  allocated to execute concurrent requests for a function automatically
scaled up or down based on the incoming workload. Since containers are lightweight to provision,
such scaling can be done in a near instantaneous fashion allowing for quick reaction to incoming
bursts of requests.

Serverless platforms have recently been the focus of many research efforts \cite{castro2017serverless,
hellerstein2018serverless, wang2018peeking}. In addition, all major commercial cloud providers have
serverless cloud offerings\cite{AWS-Lambda, Azure-Functions, Google-Cloud-Functions}. However, a
less studied form of this emerging paradigm is serverless computing at the edge. Edge computing
involves deploying computational resources at the edge of the network close to end-users, making it
particularly well-suited for latency-sensitive or bandwidth-intensive applications
\cite{satyanarayanan2017emergence, shi2016edge}.
In many cases, the computational needs of such applications may be bursty and processing may
involve short-lived computations, making a serverless model attractive for edge computing
\cite{aske2018supporting}. Consider the following motivating example of IoT data processing in the
edge.


\noindent{\textbf{\em Example 1:}} Consider an IoT device such as a motion-activated smart camera
that captures and streams video frames only when it senses motion, yielding a bursty data stream.
This video data needs to be analyzed using some object detection and recognition deep learning
inference model \cite{kang2019deeprt,wang2018pelee}. The inference needs to be performed in near
real-time to generate alerts in case of the detection of a suspicious object in the video. Such
application workload can benefit from both edge computing and serverless computing; Since the video
data has high bandwidth needs, especially for high-definition (HD) cameras, processing this data at
the edge saves network resources compared to sending such high-volume data to a distant cloud.
Further, since video data is not continuous and is only generated upon detecting motion, the
serverless paradigm that allocates resources to a function on demand is better than a long-running
edge application that stays idle  between requests wasting resources. Since edge resources are
constrained, serverless computing makes more efficient use of the scarce edge resources over a
persistent edge resource allocation. These observations motivate our focus on serverless edge
computing in this paper.

Serverless computing at the edge raises a different set of research challenges compared to
traditional serverless in the cloud. First, while the cloud can offer the illusion of ``infinite
capacity'' to auto-scale serverless functions, edge clusters are more resource-constrained by virtue
of having fewer servers imposing limits on auto-scaling. Hence, serverless edge platforms may
occasionally face resource pressure when multiple hosted functions from different users see
simultaneous bursts. Second, since serverless computations are supposed to be transient and
short-lived, typically, cloud platforms impose hard time limits on the maximum execution time
allocated to each function invocation-- the computation is terminated if it doesn't finish execution
within a limit. Commercial cloud platforms have typically imposed a time limit of up to 60 seconds
for each function invocation, although more recently, they have begun to raise this limit to a few
hundreds of seconds accommodating a broader set of serverless workloads \cite{AWS-Lambda-timeout}.
Likewise, edge servers will need to judiciously allocate scarce resources to latency-sensitive
computations such that they can finish execution before the revocation deadline.

To address the above challenges, this paper presents \emph{LaSS}, a platform for managing
\emph{La}tency \emph{S}ensitive \emph{S}erverless computations. We address the challenges described
above by designing model-driven resource management algorithms for latency-sensitive serverless
computations in edge clusters. In designing, implementing and evaluating our system, we make the
following contributions:
\begin{itemize}
    \item We present a principled approach for allocating edge resources to latency-sensitive
          serverless functions using queuing models so as to meet their SLO deadlines. We use our
          models to design an algorithm to dynamically scale the resource allocation of each
          function up or down based on time-varying workload demands.

    \item In the absence of resource pressure, our model-driven approach dynamically allocates the
          necessary resources to each function based on its observed workload. When the system
          experiences resource pressure or overloads, LaSS uses a weighted fair-share resource
          allocation approach that guarantees a minimum fair share of the edge cluster resources to
          each function in proportion to its weight. We also present resource reclamation policies
          based on container termination and container deflation to reclaim resources from
          over-allocated functions that are using more than their fair share and allocate them to
          under-provisioned functions.

    \item We implement a full prototype of our model-driven approach and fair-share reclamation
          policy on a serverless edge cluster based on the open-source OpenWhisk platform.

    \item We conduct a detailed experimental evaluation of our system using a diverse set of
          serverless workloads including multiple edge-based machine learning inference models for
          bursty video processing, Geofencing, image resizing, and malicious file detection. Our
          results show the ability of our system to quickly react to fluctuations in the workload
          seen by functions (within tens of milliseconds when load increases by 10\% and within
          hundreds of milliseconds when load increases by 100\%), adjusting their container
          allocations in order to meet their SLO deadlines. Our results also show that our deflation
          and termination-based reclamation policies enables LaSS to operate with fair-share
          allocation guarantees in overload scenarios where vanilla OpenWhisk will result in
          cascading failures in the workloads. We further show that using deflation for resource
          reclamation improved utilization by about 6\% in highly packed edge clusters compared to
          using the termination policy.
\end{itemize}

\section{Background}
\label{sec:background}

In this section, we discuss background on serverless and edge computing, as well as our problem
statement.

\subsection{Serverless Computing}
A serverless computing platform offers a \emph{Function-as-a-Service (FaaS)} abstraction where
application code is deployed at the granularity of a function and is invoked upon an event (e.g.,
request arrival or new data produced). A key advantage of the FaaS abstraction is that application
owners do not need to worry about server deployment considerations, such as how many servers to
allocate to an application or how to auto-scale the server allocation up or down. Serverless
platforms offer built-in elasticity dynamically scaling container resources allocated to a function
on-demand. AWS Lambda \cite{AWS-Lambda}, Azure Functions \cite{Azure-Functions}, and Google's cloud
Functions \cite{Google-Cloud-Functions} are  examples of FaaS offerings from major cloud providers.

There are three important characteristics of serverless functions that are relevant to our work.
First, our work assumes that a serverless function is simply a piece of code written in any language
of choice that is capable of executing inside an OS container. This feature is already supported by
most major serverless platforms \cite{Azure-Lambda-containers, Azure-Functions-custom-handlers,
OpenWhisk-blackbox-actions}. Beyond the ability to observe the container's behavior from the outside
and control the resource allocation to a container, the platform does not have any specific
knowledge of the function itself.

Second, since serverless functions are supposed to be short-lived computations, FaaS platform impose
a hard time limit on the execution of each function. The computation is terminated if it does not
complete execution within this limit. Cloud serverless platforms, for instance, have enforced
execution time limits of between 60s to up to 900s to allow a single function to process longer
input streams \cite{AWS-Lambda-timeout}. In our case, edge applications are also assumed to have
latency-sensitive needs. Hence, from a resource management standpoint, we abstract both of these
constraints using an SLO deadline---the platform needs to ensure that a high percentile of the
requests (e.g., 99\%) complete by their SLO deadline. In practice, the SLO  deadline is determined
based on the latency requirements of the application and the hard limit imposed by the serverless
platform.

Third, function requests are assumed to arrive asynchronously at a certain stochastic rate, and upon
a new arrival, the control node determines whether to forward the request to an existing container
for that function (if one exists) or to spin up a new container (via auto-scaling).

From a platform standpoint, we assume that the serverless cluster comprises a \emph{dispatcher} node
where requests or events arrive, a \emph{control} node that make resource allocation decisions, and
a group of \emph{worker} nodes each of which run one or more containers that execute functions.
Popular open-source FaaS platforms such as OpenWhisk \cite{baldini2017serverless} and Kubeless
\cite{kubeless} employ such an architecture.

\subsection{Edge computing}
Edge computing has emerged as a complement to cloud computing for running latency- and
bandwidth-sensitive applications in close proximity to end-users and their devices
\cite{shi2016edge,satyanarayanan2017emergence}. Edge computing is particularly well-suited for
processing data generated by IoT devices in domains such as smart homes, mobile health and smart
transportation. Some researchers have even argued that video analytics over IoT data is the ``killer
application'' for  edge computing \cite{ananthanarayanan2017real}. A large number of these workloads
will depend on ``AI at the edge'', where the edge application runs, for example, a deep learning
(DL) model to perform inference over IoT data (e.g., a smart traffic intersection camera).
In our work, edge processing is assumed to be performed using serverless functions, each of which
processes a small amount of the incoming data stream.

In this paper, we focus on the problem of scheduling and resource allocation for latency-sensitive
serverless functions inside a single resource-constrained edge cluster. We do not consider the
problem of job distribution across multiple edge clusters or the problem of job scheduling over
heterogeneous edge devices. On the other hand, although we mainly discuss running serverless
functions at the edge in this paper, most of the algorithms and techniques can be applied to
cloud-based FaaS systems as well.

\begin{figure}
    \centering
    \begin{minipage}[t]{\linewidth}
        \centering
        \includegraphics[width=\textwidth]{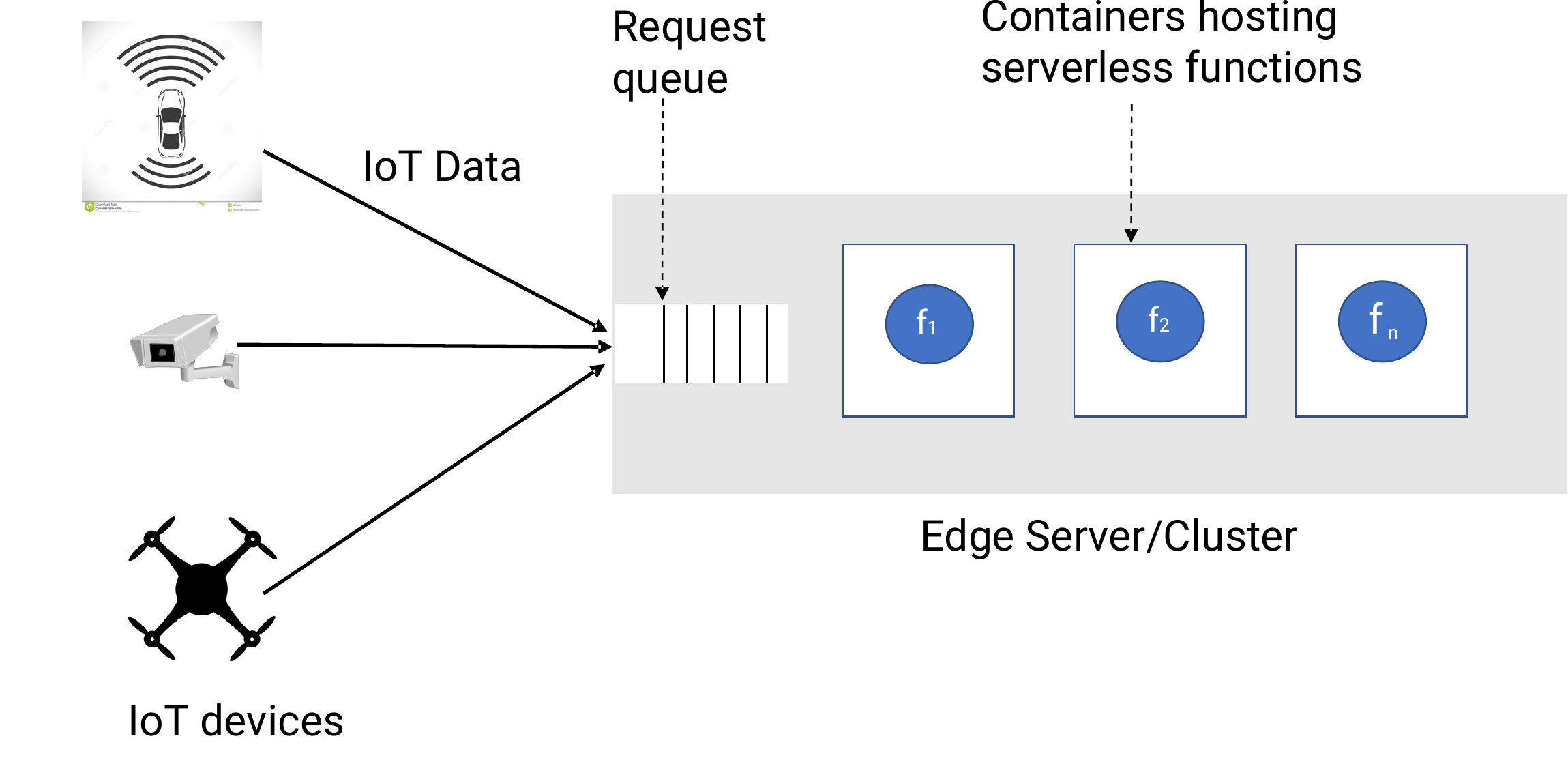}
        \vspace{-7 mm}
        \caption{An Edge cluster running serverless computations that process dynamic IoT data.}
        \label{fig:edge}
    \end{minipage}
\end{figure}

Figure~\ref{fig:edge} depicts the serverless edge platform considered in this paper. Edge clusters
are resource-constrained. So long as end-users or devices produce  data bursts at different times,
the cluster has enough resources to scale each function to the desired number of containers to
process this incoming burst in parallel (and meet SLO requirements). However, whenever multiple
functions belonging to different users see concurrent workload, the cluster may see a temporary
overload  where it lacks capacity to service the aggregate workload. In this case, it will need to
carefully and fairly allocate resources across competing serverless functions. Doing so may require
reclaiming resources from functions that are using more than their fair share and reassigning them
to under-provisioned functions. Thus, the edge cluster should offer auto-scaling  to  serverless
computations in the absence of resource pressure falling back to fair share allocation  during
periods of high loads.

\subsection{Problem Statement}
Consider an edge cluster with $N$ edge servers and a control node. Let $f_1.....f_k$ denote $k$
functions that are hosted on this cluster. Let $\lambda_i$ denote the mean inter-arrival time of
requests or data  for function $i$.
Let $d_i$ denote the the SLO deadline for function $i$, where the SLO requires a certain high
percentile of requests (e.g., 95\% of the requests) to complete execution by the deadline. Let
$\mu_i$ denote the mean execution time of function $i$. To finish by the deadline $d_i$, we require
that $\mu_i+q_i \leq d_i$  for a high percentage of the requests, where $q_i$ is the time spent
waiting in the edge system for a container.

Under normal load, the control node should allocate enough containers to each function so that
incoming events can finish by the deadline $d_i$ by ensuring the waiting time to be scheduled onto a
``container’’ is smaller than $(d_i-\mu_i)$. In the presence of resource pressure or overload, SLAs
may be temporarily violated but such violations should be kept to a minimum by ensuring that each
function gets a minimum, fair-share allocation. This is done by assigning a weight $w_i$ to each
function that determines its allocation relative to other functions.

Our goal is to design an edge serverless system that dynamically allocates $c_i$ containers to
function $f_i$ such that the response time of incoming requests, i.e., $\mu_i+q_i$, is  less than
the SLO deadline $d_i$ for a high percentile of the requests. As the workload  $\lambda_i$
fluctuates over time, the number of allocated containers  $c_i$ should be varied so as to ensure
adequate capacity is allocated to function $f_i$ to meet its SLO deadlines. During overload scenario
(i.e., $\sum_i c_i \ge$ total cluster capacity), each function $f_i$ should be allocated a minimum
guaranteed container capacity that is in proportion to its weight $w_i$ (by reassigning container
capacity from functions that are using more than their fair share allocation to under-allocated
functions).

\section{Model Driven Resource Allocation}
\label{sec:model}

We next present our model-driven approach to determine the container capacity that should be
allocated to each function's incoming workload to meet its SLO deadline. We also present an
algorithm that uses these model estimates to scale the number of containers allocated to each
function and map them to nodes in the cluster.

\subsection{Queuing Theoretic Model}

To estimate the number of containers required for each function $f_i$, we assume that time is
partitioned into epochs. Let $\lambda_i$ denote the observed arrival rate of requests to function
$f_i$ in the previous epoch.
Let us assume the function $f_i$ needs $c_i$ containers to service this workload $\lambda_i$ while
meeting the SLO deadline $d_i$. Our goal is to determine an appropriate value of $c_i$.

Initially, let us assume that all $c_i$ containers are homogeneous---they have identical CPU and
memory allocations. Since containers allocated to a function have the same resource capacity, we can
model them using the same service rate $\mu_i$ for processing requests. Hence, service time
(execution time) of a request is given by $s_i = 1/\mu_i$. If we assume that request arrivals are
Poisson and that services times are exponential, we can model this system using a M/M/c/FCFS queuing
system with $c$ queuing servers to process incoming requests. In our case, each queuing ``server''
is a container that processes requests rather than an actual server. For function $f_i$, $c=c_i$,
since there are $c_i$ concurrent containers, each of which processes incoming requests to this
functions in an FCFS manner.

While the arrival process of requests for a serverless function may not be Poisson in the long run,
we argue that on shorter time scales they can be approximated as a Poisson process. For instance, in
the motion triggered camera example the request rate is approximately a Poisson process over a short
time window after the camera is triggered. Therefore, we can use a moving window approach to
estimate the current request rate of each function, and every time the algorithm runs it uses the
arrival rate calculated over the previous epoch as the input arrival rate

Queuing analysis of an M/M/c system is well known in the literature and yields a closed-form
equation the probability $P_n$ that an incoming requests sees $n$ other requests in the system. The
steady state probability $P_n$ for an M/M/c system is given as~\cite{harchol2013performance}
\begin{equation}\label{eq:violation}
    P_n= \begin{cases}
        \frac{r^n}{n!} P_0       & 0\leq n \leq c. \\
        \frac{r^n}{c^{n-c}c!}P_0 & n\geq c.
    \end{cases}
\end{equation}
where,
\begin{equation}
    P_0= \Bigg(\frac{r^c}{c!(1-\rho)} + \sum_{n=0}^{c-1}\frac{r^n}{n!}\Bigg)^{-1} \label{eq:3}
\end{equation}
The term $r$ is shorthand for $r=\lambda_i/\mu_i$, $c=c_i$, and $\rho$ denotes the utilization of
the system given by $\rho=\lambda_i/c_i\mu_i$. We can use these steady-state probabilities of the
system to derive the waiting time distribution and bounds on a higher percentile of the waiting time
(which can then be used to meet the SLO deadline $d_i$).

Specifically, let $n$ denote the number of requests seen by an incoming request for the function.
Since there are $c_i$ containers, if $n<c_i$ there is at least one idle container and the request
does not see any queuing delay and can immediately be serviced by an idle container. If $n\geq c_i$
all $c_i$ containers are serving existing requests and the request must wait until a container
becomes idle. Since requests are processed at mean rate $\mu_i$ and $c_i/\mu_i$ requests get
processed per unit time, the expected waiting time is $(n_i-c_i+1)/\mu_i c_i$ Let $t$ denote the
upper bound on the waiting time (we discuss how to derive $t$ using the deadline below). therefore,
\begin{equation}
    \frac{n-c_i+1}{\mu_i c_i} \leq t \label{eq:4}
\end{equation}
This yields $n\leq tc_i \mu_i+c_i-1$, which implies that the expected waiting time is less than t
when the number of requests $n$ is less than or equal $\mu_i c_t +c_i-1$. The probability of seeing
no more than $\mu_i c_i t +c_i-1$ customers can be derived from the steady state probability as
$\sum_{n=0}^{c_i} P_n$. Hence the probability of bounding the waiting time $Q$ by $t$ is
\begin{equation}
    P(Q \leq t) = \sum_{n=0}^{c_i} P_n
\end{equation}

In our case, the SLO deadline $d_i$ must be satisfied for a high percentile of the requests (say the
99\textsuperscript{th} percentile). In the worst case, a request sees a high service time (e.g.
99\textsuperscript{th} percentile of service time distribution) and a high wait time
(99\textsuperscript{th} percentile) and the sum of the two should still be less than $d$. Hence, we
can set $t_{p99}=d-1/\mu_{p99}$ where $t_{p99}$ denotes the 99\textsuperscript{th} percentile of the
waiting time distribution and $1/\mu_{p99}$ is the 99\textsuperscript{th} percentile of the service
time. We can then substitute $t_{p99}$ into equations \ref{eq:3} and \ref{eq:4} iteratively to find
the smallest $t$ such that the right-hand sum equals the 0.99. Algorithm~\ref{alg1} denotes this
iterative procedure. We note that this calculation converges rapidly to the correct number of
servers. Our queuing model allows us to compute the number of containers $c_i$ needed to service the
observed incoming workload $\lambda_i$ while meeting deadline SLO $d_i$ for a high percentile of the
requests.
\begin{algorithm}
    \SetAlgoLined
    \KwData{The request arrival rate $\lambda$, the service rate $\mu$, $t_{p99}$}
    \KwResult {The number of containers required $c$}
    \BlankLine
    $c \leftarrow$ number of containers in the system \;
    \While{$P \leq 0.99$}{
        $c \leftarrow c + 1$\;
        $L \leftarrow \floor{tc\mu + c -1}$\;
        $P = \sum_{n = 0}^ L p_n$\;
    }
    \Return c
    \BlankLine
    \caption{Iterative algorithm for finding $c$}
    \label{alg1}
\end{algorithm}

\vspace{-5 mm}
\subsection{Modeling Heterogeneous Containers}
\label{sec:heterogeneous_model}

Our above model assumed all containers, $c_i$, allocated to function $f_i$ have homogeneous CPU and
memory allocations. However, due to the resource reclamation techniques discussed in
Section~\ref{sec:fairness}, the allocation of the $c_i$ containers may not be identical, leading to
different service rates. To model this scenario, let us assume that $\mu_i^j$ denotes the service
rate of the $j^{th}$ container, where $1\leq j \leq c_i$, without loss of generality, assume
$\mu_i^1\leq \dotsb\leq \mu_i^{c_i}$ We can compute the steady state probabilities $P_n$ of seeing
$n$ requests in the system based on the worst case analysis in Alves et al.~\cite{Alves2011Upper}.
The worst case analysis is based on the observation that, under heavy traffic, the probability of
having a job running on the slowest container(s) is higher than having a job running on the fastest
one, which can then be used to derive a bound on the waiting time less than or equal to $t_{p99}$.
As this is worst-case analysis, we assume, like Alves et al., that the scheduler will always make
the worst decision, scheduling jobs first on the slowest container(s), and last on the fastest
container(s). Alves et al. show that the upper bound probability $P_n$ that an incoming request sees
$n$ requests in the system when $n<c_i$ is:
\begin{equation}
    P_n=P_0 \frac{\lambda^n}{\prod_{k=1}^n (\sum_{j=1}^k \mu_j)}, \label{eq:homog}
\end{equation}
and in case of overload when $n>c_i$ is:
\begin{equation}
    P_n=P_0 \prod_{k=1}^c\bigg(\frac{\lambda}{\sum_{k=1}^c \mu_k}\bigg) \bigg[\prod_{k=c+1}^n \bigg(\frac{\lambda}{\sum_{k=1}^c \mu_c}\bigg) \bigg].   \label{eq:hetro}
\end{equation}
We are mostly interested in the case when $n>c_i$, which typically results in multiple functions
being deflated, as we describe in the next section, and thus having wide differences in service
rates. LaSS thus implements Equation~\ref{eq:hetro} in an iterative algorithm similar to
Algorithm~\ref{alg1} described previously.
As we will show experimentally in \S \ref{sec:eval}, our simple iterative algorithm scales well to
very large number of serverless functions, making the use of queuing models practical in serverless
edge platforms. Finally, we note that this worst-case analysis assumes that the scheduler uses a
heuristic that picks the absolute slowest container(s) first, thus adding to the queuing and delays,
instead of, e.g., using the fastest containers first.

\subsection{Container Allocation Algorithm}
\label{subsec:allocation_algorithm}
The above model yields the number of containers $c_i$ that should be allocated to each function
$f_i$   to handle the expected workload  $\lambda_i$.  To implement our model-driven approach, the
control node in our serverless edge cluster monitors the incoming workload $\lambda_i$ for each
function within each epoch. At the end of an epoch, it recomputes the number of containers $c_i$ to
be allocated to each function $f_i$ hosted within the cluster based on the above models. In order to
be responsive to bursts of requests, epochs are relatively short in our system (e.g., tens
of seconds to a minute). The observed request rate  in each epoch yields a time series of per-epoch
observations that is subjected to an exponential weighted moving average (EWMA) with a high weight
given to the most recent epoch. The resulting workload $\hat{\lambda_i}$ for a function $f_i$, the
observed service time $\mu_i$, and the SLO deadline $d_i$ are used as inputs to our model to compute
a container allocation $c_i^{new}$ for each function.

If the current container allocation $c_i^{current}$ is greater than $c_i^{new}$, the function is
over-provisioned and ($c_i^{current}-c_i^{new}$)  containers with the lowest resource allocations
are marked for termination. On the other hand, if $c_i^{new}>c_i^{current}$, the function needs more
resources, and the controller node needs to start $c_i^{new}-c_i^{current}$ additional containers to
handle the workload. Hence, the control node first finds a cluster node with enough spare capacity
or finds a number of nodes that can collectively host $c_i^{new}-c_i^{current}$ new containers. It
then signals the invoker on each node to start these additional containers. If insufficient idle
resources are available, any container marked for termination for over-provisioned functions is
actively terminated, and those resources are reallocated to under-provisioned functions. Note that
in absence of resource pressure, the cluster is guaranteed to find adequate resource capacity to run
$c_i^{new}$ containers for each function. In the  overload scenario, Equation~\ref{eq:hetro} is
used. Finally, note that containers marked for terminations are reclaimed in a lazy fashion and only
when needed. Doing so allows them to be reused if the load increases again.

\section{Fair-share Resource Allocation and Reclamation}
\label{sec:fairness}

The previous section assumed an absence of resource pressure where each function could receive its
desired container allocation as computed by our model.
However, edge clusters are more resource-constrained than a centralized cloud in terms of their
server capacity, and occasionally an edge cluster may face resource pressure where the total
resource capacity needed to host all containers for all functions exceeds the cluster capacity.  In
this section, we describe techniques for handling resource pressure based on (i) fair share
allocations and (ii) resource reclamation.

\subsection{Fair Share Resource Allocation}

To deal with scenarios when the edge cluster sees an overload, our system \emph{guarantees} a
minimum resource allocation to each function. In the absence of  resource pressure, a function's
container allocation is allowed to {\em exceed} this minimum share since our goal is to
allocate sufficient capacity to meet SLO deadlines of latency-sensitive requests. An overload is set
to occur when the aggregate container capacity across all functions ($\sum_i c^{new}_i$) exceeds
the total cluster  capacity.  Hence, when an overload occurs, any function that is allocated more
container capacity than its guaranteed minimum share is reduced to no less than minimum allocation,
and functions that have allocations under their guaranteed  limit are given additional resources up
to this limit.

To ensure fairness in capacity allocation during overloads, each function is assigned a weight
$\omega_i$ by its owner. The guaranteed minimum fair share allocation of each function is
proportional to its weight -- each function is guaranteed $\omega_i/\sum_j \omega_j$ fraction of
cluster resources. Doing so ensures a minimum rate of execution for a function under overload and
avoids starvation or unfairness where a function with greater workloads takes an unfair share of
cluster capacity.

Our system then uses the following algorithm to determine the container allocations for each
function. It first uses the queuing models from the previous section to determine the desired
container allocation $c_i^{new}$ for each function. In the presence of overload $\sum_i c_i^{new}>C$
where C denotes the total cluster capacity in terms of number of containers.
Let $c_i^{guar}$ denote the guaranteed minimum share for function $i$ (again in terms of
containers), where
\begin{equation}
    c_i^{guar}=\floor*{\sum_j\frac{\omega_i}{\omega_j} \cdot C}.
\end{equation}
If $c_i^{new}\leq c_i^{guar}$, then resource demand for function $i$ is equal to or
below its guaranteed minimum share, and hence is allocated its desired capacity $c_i^{new}$
as computed by our model. Such functions are ``well behaved’’ and do not require any reduction in
the desired container capacities. Thus, for well-behaved functions $c_i^{adj} = c_i^{new}$, where
$c_i^{adj}$ denotes the adjusted allocation of function $f_i$.

Let $ \hat{C}=C- \sum_k c_k^{new}$ denote the remaining capacity in the cluster after allocating
capacity to all well-behaved functions $k$. The remaining functions are those where
$c_i^{new}>c_i^{guar}$, i.e., their desired capacity exceeds their guaranteed minimum share. Each
such function in then allocated the remaining cluster capacity $C'$ in proportion to their weight.
\begin{equation}
    c_i^{adj}=\floor*{\sum_m\frac{\omega_i}{\omega_m}\hat{C}}
    \label{eq:ci}
\end{equation}
where $m$ denotes the set of overloaded functions.

The above algorithm guarantees that all overloaded functions receive at least their guaranteed fair
share allocation as shown in the following lemmas.

\noindent\textbf{Lemma 1:} In the scenario where all functions are overloaded,
$c_i^{new}>c_i^{guar}$, our algorithm allocates the guaranteed fair share $c_i^{guar}$ to each.

\noindent\textbf{Proof.} Since {\em all} functions are overloaded, there are no well-behaved
functions in the system. Since the set $k$ of well-behaved functions is empty, the expression $
\hat{C}=C- \sum_k c_k^{new}$ reduces to $\hat{C}=C$. By substituting for $\hat{C}=C$ in Equation
\ref{eq:ci}, it follows that each function is allocated exactly its guaranteed share
$\floor*{\sum_m\frac{\omega_i}{\omega_m}\hat{C}}$

\noindent \textbf{Lemma 2:} In the case where only some functions are overloaded, each such function
receives no less than its guaranteed minimum share  $c_i^{guar}$.

\noindent\textbf{Proof.} Since each well-behaved function is allocated  its desired allocation
$c_i^{new}$ which is less than or equal to its guaranteed share $c_i^{gaur}$, it following that for
all well-behaved functions $\sum_k c_k^{new} \leq \sum_k c_k^{gaur}$, where $k$ denotes the set of
well behaved functions. Since $ \hat{C}=C- \sum_k C_k^{new}$, it follows that the remaining capacity
$\hat{C}$ is greater than or at least equal to to the guaranteed share  of the remaining overloaded
functions. That is $\hat{C}\geq \sum_m c_m^{guar}$, where $m$ is the set of overloaded functions.
This remaining capacity is assigned in proportion to weights of the overloaded functions as per
Equation \ref{eq:ci}, giving each function a share that is greater than or equal to its fair share
allocation.

\subsection{Resource Reclamation Algorithms}
Once our algorithm determines the new allocations $c_i^{adj}$ for each function, our system needs to
adjust the current allocations to these new values. This typically involves reclaiming resources
from overloaded functions and reducing their allocation to a $c_i^{adj}$, which is at least the
guaranteed share. The reclaimed capacity is given to any function that requires a capacity
increase---where the new allocation $c_i^{adj}$ is higher then the current allocation
$c_i^{current}$.

There are two mechanisms to reclaim resources from over-allocated functions: container termination
and container deflation.
\begin{enumerate}
    \item \textbf{Termination.} Termination simply involves gracefully shutting down the Container
          and reclaiming its CPU and memory resources. Unlike lazy termination where resources were
          reclaimed in a lazy manner in the absence of resource pressure, container termination is
          immediate during overload situations.
    \item \textbf{Deflation.}  Resource deflation is a recently proposed approach for reclaiming
          resources by reducing the CPU and memory allocation of a virtual machine. The
          deflation approach proposed in \cite{sharma2019resource,hpdc20} is based on the
          observation that virtual machines have slack in their resource allocation (since their
          allocated resources may not be 100\% utilized by the application code executing inside
          them) and reclaiming a fraction of the VM's allocated resource can be done without a
          proportionate degradation in application performance. The deflation study in \cite{hpdc20}
          has analyzed millions of production VMs in the Azure cloud and showed that typical slack
          can be up to 50\%.

          We adopt the VM deflation idea to containers. Since containers are allocated a certain CPU
          and RAM allocation by the OS, this allocation can be dynamically changed using system
          calls to implement the notion of deflation for OS containers. Since our containers are
          latency-sensitive functions, such deflation must not be too aggressive to avoid
          performance degradation that causes SLO deadline violations. We do limit the impact on
          performance by specifying a maximum threshold  on the fraction of  resources reclaimed
          from a deflated container. While this threshold is application-dependent, in our current
          implementation we set this value conservatively (e.g., $\tau= 30\%$) to limit the
          performance impact.
\end{enumerate}
These two options for resource reclamation yields two different policies to reclaim resources from
over-allocated functions during an overload. The termination-based reclamation policy examines all
over-allocated functions where we $c_i^{adj}>c_i^{current}$ and terminates $c_i^{adj}-c_i^{current}$
containers for each such function. Using the freed up resources, it then reallocates those resources
to all under-allocated functions by increasing their allocation to the computed value $c_i^{adj}$.

The deflation-based reclamation policy examines over-provisioned functions where $c_i^{new} >
c_i^{current}$ and iteratively deflates the $c_i^{current}$ containers of each such function by up
to a threshold amount, in small increments. After each incremental iteration, it checks if the total
CPU capacity allocated to the deflated $c_i^{current}$ containers equals the total CPU allocation of
the desired $c_i^{adj}$ non-deflated containers. If not, it iteratively deflates all $c_i^{current}$
container by another increment,  up to a threshold to $\tau$, until sufficient resources have been
reclaimed (i.e., the aggregate capacity of the deflated containers equals the total capacity of
$c_i^{adj}$ non-deflated ones.

If sufficient CPU capacity is not reclaimed even after deflating each container by the maximum
value defined by $\tau$, some containers are terminated until the aggregate CPU allocation of the
deflated containers equal that of $c_i^{new}$ non-deflated ones. A key advantage of the deflation
approach is that it allows a function to have strictly more containers than the termination-based
reclamation approach. This allows more concurrency where requests can be processed in parallel
reducing waiting time and reducing SLO deadline violations or the magnitude of the violations. As
long as deflation is done conservatively, the service time should see only a small degradation
while benefiting from higher concurrency (see \S \ref{sec:eval}).

\section{LaSS Implementation}

\begin{figure}[htb]
    \centering
    \begin{subfigure}{\linewidth}
        \centering
        \includegraphics[width=\textwidth]{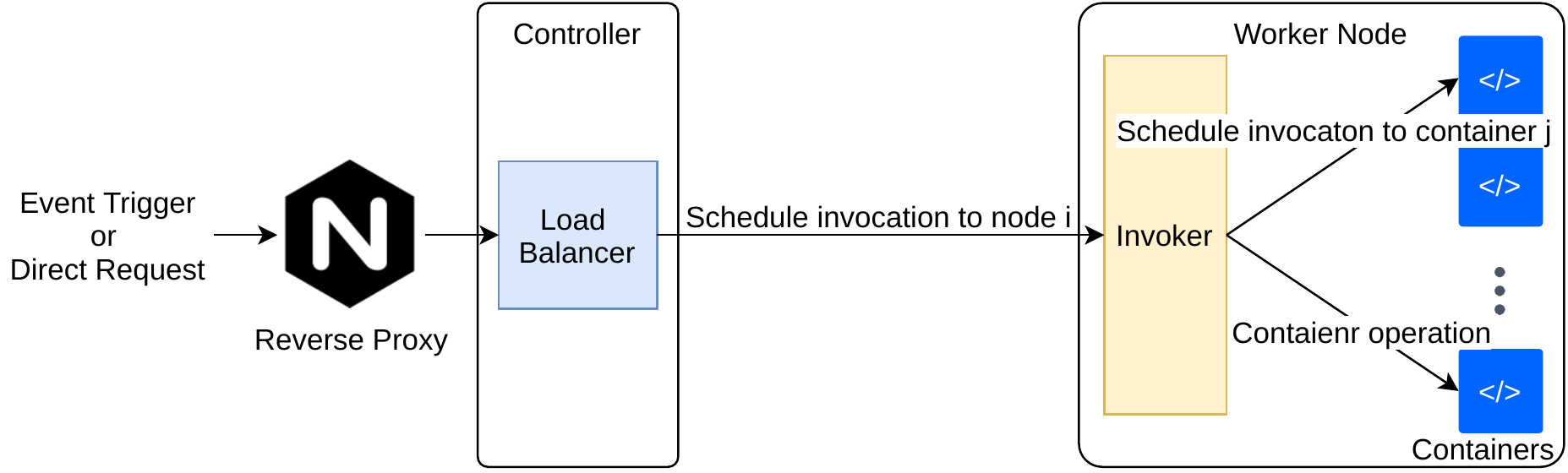}
        \caption{Architecture of OpenWhisk.}
        \label{subfig:arch-openwhisk}
    \end{subfigure}
    \begin{subfigure}{\linewidth}
        \centering
        \includegraphics[width=\textwidth]{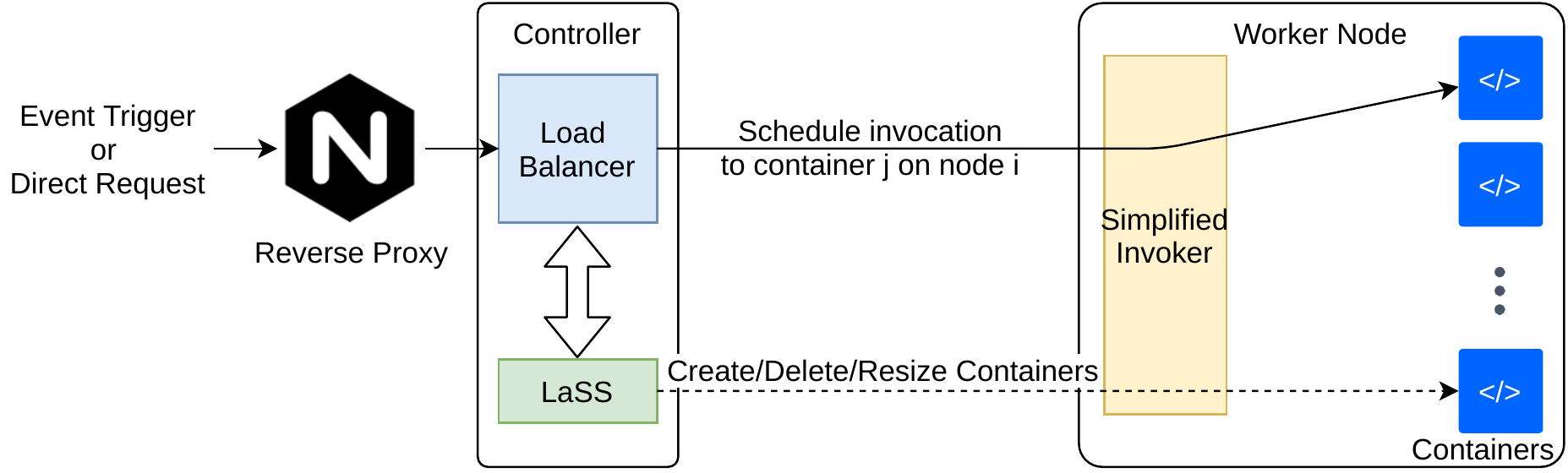}
        \caption{Architecture of the LaSS prototype based on OpenWhisk.}
        \label{subfig:arch-lass}
    \end{subfigure}
    \vspace{-3 mm}
    \caption{Architectures of OpenWhisk and the LaSS Prototype.}
\end{figure}

Our system is implemented based on of Apache OpenWhisk \cite{openwhisk}, a popular open-source
serverless framework implemented in Scala. The architecture of OpenWhisk is shown in Figure
\ref{subfig:arch-openwhisk} (components of less relevance have been omitted for clarity). In
OpenWhisk, when a function is invoked, either by event trigger or by direct request, the invocation
request is sent to a controller. The load balancer in the controller will schedule the invocation to
one of the worker nodes based on their health and load status. When the request reaches the assigned
worker node, the invoker on that node will execute the invocation request inside a container.

One major issue of OpenWhisk is that the control path and the data path are coupled, which makes it
unsuitable for running latency sensitive computations at the edge. In OpenWhisk, the controller is
in charge of deciding which invoker an invocation should be scheduled to, but the container level
decisions (e.g., whether to reuse an existing container or create a new one) are made by the
invoker. This design leads to a gap between control and information in the system---invokers make
decisions of creating/removing function instances but it only has local information; controller, on
the other hand, has more global information but does not have direct access to the containers. This
makes it impossible to provide SLO guarantee in the absence of resource pressure or to enforce
fairness during overload, since there is no way to control the exact number of containers in the
system for a particular function.

Figure~\ref{subfig:arch-lass} shows how we integrate LaSS into OpenWhisk. Our changes are mainly
localized to OpenWhisk's controller and invoker components. We separated the control path (the
dashed line) and the data path (the solid line) by adding a dedicated LaSS module in the controller.
The LaSS module has direct control over all containers in the system and can also read arrival rate
statistics from the load balancer. Using the arrival rate information our LaSS module can
periodically update the containers allocated to each function using the resource allocation
algorithm we discussed in the previous two sections, as well as the load balancer configuration.
Knowing all the containers and their size information, the load balancer uses the weighted round
robin (WRR) algorithm to directly schedule function invocation request to each individual container.
The invoker is now much simplified, it no long make any decisions on scheduling or container
operation, it only executes commands from the controller. These enhancements allow our system to
allocate the desired container capacity to meet SLOs in the normal case, while also providing
fairness guarantee under overload.

Another major change we made to OpenWhisk is to allow users to directly specify the CPU
request/requirement for a function. OpenWhisk does not allow that but it does provide the user with
the ability to set memory allocation per function when creating a function. CPU resources are then
automatically allocated in proportion to the memory. However, this is inadequate for latency
critical serverless functions since; (1) It is variable and machine dependent -- the amount of CPU
allocated depends on the physical machine and the other functions that are co-located on the same
host, and (2) Functions have different resource requirement patterns. For example, some functions
may be CPU-constrained requiring a lot of CPU, but much less memory resources. Therefore we have
added the ability for users to control both CPU and memory allocation for their functions to better
serve latency sensitive computations.

For function scheduling, we implemented a two level hierarchical scheduling tree by adding the
notion of weight to user (namespace) and actions. LaSS uses these weights to calculate the fair of
resources for each action. Our model can be extended to a hierarchical scheduling tree with
arbitrary levels.

In our implementation, OpenWhisk components are deployed using Kubernetes (to streamline the
deployment process), while the functions run in native Docker containers. This is because to
implement deflation we need to have the ability to update the resources of a container ``in place'',
i.e., to adjust the resources allocated to a running container without interrupting the running
function. Kubernetes currently does not support that, instead it will first create a new container
with the new specifications and then destroy the old one. On the other hand, although it is possible
to update the resource allocation of a Docker container during runtime, deflation of memory is
non-trivial: a container may get killed for exceeding its memory limit if the updated limit is less
than the memory already in use. Therefore, we currently only implement CPU deflation in LaSS, and
the memory allocation of a function instance will not change when its CPU allocation gets deflated.

In order to use queueing theory based models to predict the capacity needed for a latency sensitive
function, the controller needs to know the service time distribution. In the scenario where the
deflation policy is used, the controller needs to know multiple service time distributions under
different container sizes. LaSS supports two approaches for this purpose: 1) load offline profiling
results which may be measured by either the user or the service provide, and 2) use an online
learning algorithm to learn the service time distribution(s) over time.

Finally, for auto scaling, LaSS also need to estimate the arrival rate of a function. In our
prototype LaSS accomplish this by monitoring two sliding windows every 5 seconds: a 2-minute long
window and a 10-second short window. When no burst is detected, the arrival rate is calculated using
the long window, but when there is a burst, i.e., if the arrival rate in the short window is twice
as high as the arrival rate in the long window, LaSS switches to calculating the arrival rate based
on the short window. This is largely inspired by another system, Knative \cite{knative}. Note that
we don't imply this is the best way to estimate arrival rate. It can be argued that predicting
arrival rate using time series analysis or machine learning techniques may be more effective.
However that is out of the scope of this paper. We chose this implementation merely because of its
simplicity. One can also plug in any load prediction method of choice into LaSS with ease.

Our modifications to OpenWhisk were distributed across 55 files in the OpenWhisk code-base, adding
over 2300 lines of code. Source code for our LaSS prototype is available publicly at \url{https://github.com/umassos/lass-serverless}.

\section{Experimental Evaluation}
\label{sec:eval}

In this section, we conduct a detailed experimental evaluation of our approach presenting our
experimental methodology, and results.

\subsection{Experimental Setup}

\noindent \textbf{Hardware.} Our experiments are conducted on a small edge cluster of 3 nodes with
each node comprising a 4-core Intel Xeon E5 processor with 16GB RAM and 10 Gbps Ethernet. All nodes
run Ubuntu 18.04 LTS Server and our OpenWhisk-based prototype. We chose this relative small cluster
setup because it is easier to create resource contention, which helps us evaluate the resource
reclamation algorithms and policies. However, it is worth noting that our system can scale to much
larger clusters, as we will demonstrate later in \S \ref{subsec:model_scalability}.

\noindent \textbf{Functions.} We use six realistic serverless functions chosen from various edge
computing applications in addition to a micro-benchmark functions. The programming languages used
for implementation and size of a standard container of each function is show in Table
\ref{table:functions}. The details of all the functions are described below:

\begin{itemize}
    \item First we implemented a configurable micro-benchmark serverless function that performs
          mathematical computations. We have the ability to control the amount of CPU cycles
          consumed by each invocation by passing a parameter with the invocation request.

    \item We also chose three deep neural network (DNN) inference models: MobileNet v2
          \cite{sandler2018mobilenetv2}, ShuffleNet v2 \cite{zhang2018shufflenet}, and SqueezeNet
          \cite{iandola2016squeezenet}. These DNN inference models are all designed for lightweight
          image classification and object detection, therefore commonly used by image/video
          processing applications in edge environments. These functions are set to emulate a motion
          activated camera that sends a burst of image data upon detecting motion. The image data is
          fed to one of the three DNN models for inference and object detection. Each ML inference
          task on an image is a separate invocation of the serverless function running one of these
          DNN models. We used the reference implementations from the \texttt{torchvision} package
          \cite{torchvision} and wrapped them in Docker as OpenWhisk blackbox functions.

    \item Our next function is BinaryAlert \cite{binaryalert}, an open-source serverless real-time
          framework for detecting malicious files. It was originally designed to run on AWS Lambda.
          We have adapted BinaryAlert to run on OpenWhisk for our experiments.

    \item Geofencing enables users to create virtual perimeters for some objects, e.g., drones,
          notifying the application owner when the objects come-in or go-out of the geofence. This
          application lends itself well to the serverless paradigm. Our sixth workload is a
          geofencing service that triggers an alert if an object leaves the virtual perimeter.

    \item Our final function is an image resizing service where images are sent for resizing. This
          is a very common use case for serverless computing.
\end{itemize}

\begin{table}
    \centering
    \begin{tabular}{p{0.3\linewidth} | p{2cm} | p{0.3\linewidth}}
        \hline
        Function        & Programming Language(s) & Standard Size     \\
        \hline
        Micro-benchmark & Python                  & 0.4 vCPU + 256 MB \\
        MobileNet v2    & Python                  & 2 vCPU + 1024 MB  \\
        ShuffleNet v2   & Python                  & 1 vCPU + 512 MB   \\
        SqueezeNet      & Python                  & 1 vCPU + 512 MB   \\
        BinaryAlert     & Python                  & 0.5 vCPU + 256 MB \\
        GeoFence        & JavaScript              & 0.3 vCPU + 128 MB \\
        Image Resizer   & JavaScript, WASM (C)    & 0.8 vCPU + 256 MB \\
        \hline
    \end{tabular}
    \caption{Functions used in our evaluation experiments}
    \label{table:functions}
    \vspace{-4mm}
\end{table}

In order to simulate the diverse set of functions that may be running simultaneously in an edge
cluster, we run a mix of deep learning functions along with regular functions in each experiment.
For all our experiments, we use a default SLO deadline of 100 ms for our serverless computation
unless specified otherwise and require that $95^\text{th}$ of waiting time should be under this
deadline (i.e., 95\% of requests should start being processed by one of the function instances
within 100 ms) unless specified otherwise.

\noindent \textbf{Workload.} For each of these serverless functions, we implemented a configurable
IoT workload generator. We use the IoT workload generator to generate invocation requests and send
the requests to the edge cluster for processing. The generator can adjust the arrival rate of
requests using on one of the following approaches:

\begin{itemize}
    \item Static. The requests are generated at a static arrival rate $\lambda$.
    \item Discrete change. The arrival rate changes at certain discrete time instants and remains
          constant in between.
    \item Continuous change. The arrival rate is adjusted after each request.
\end{itemize}

\noindent \textbf{Azure Traces.} In addition to synthetic traces, we also used traces from the Azure
Public Dataset \cite{shahrad2020serverless} in our evaluation. Since the function invocation traces
in the Azure dataset are aggregated per minute, we made our load generator work in the discrete
change mode that adjusts the arrival rate each minute when using these traces. Results using these
traces are discussed in Section~\ref{sec:AzureE}.

\subsection{Model Validation}
\label{subsec:model-validation}

Our first experiments evaluate the efficacy of our queueing models for homogeneous/heterogeneous
containers by validating their predictions experimentally.

\subsubsection{Homogeneous Containers}

In this experiment, we use our micro-benchmark serverless function and configure it with two
different service times (100 ms service time corresponding to $\mu = 10$ req/s, and 200 ms service
time corresponding to $\mu = 5$ req/s). We also tested under two different SLO deadlines (100 ms and
200 ms). We vary the arrival rate from 10 to 50 in steps of 10 and compute the container allocation
$c$ using our model. The function is then configured with $c$ containers and we empirically measure
the waiting time seen by requests and compute the $95^\text{th}$ percentile of waiting times. Each
experiment is run for 30 minutes.

\begin{figure*}[htb]
    \vspace{-0.5cm}
    \centering
    \begin{subfigure}{0.24\linewidth}
        \centering
        \includegraphics[width=\linewidth]{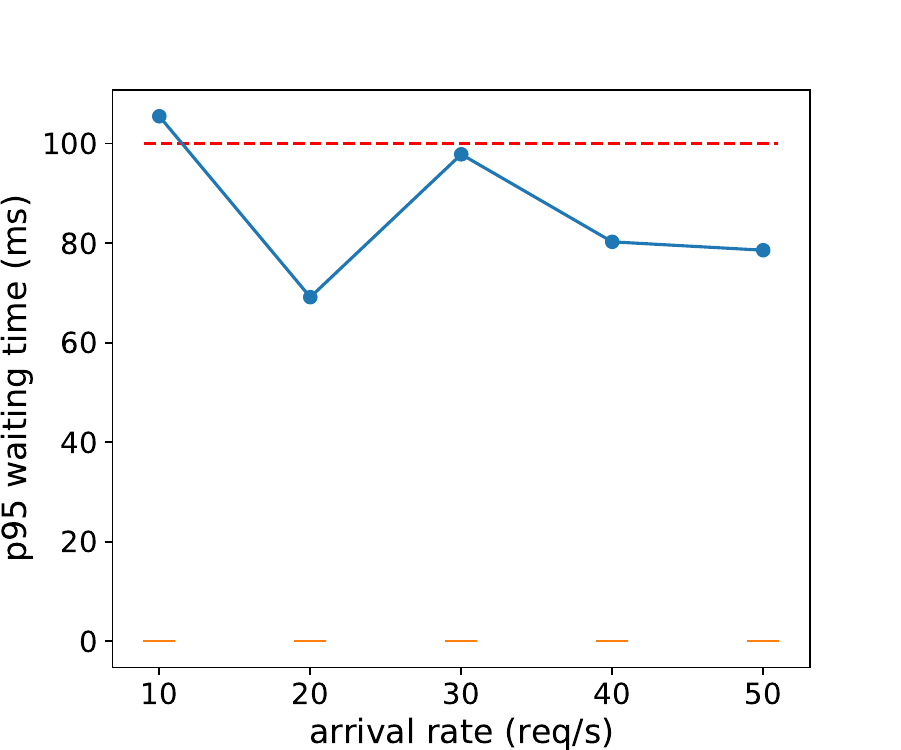}
        \subcaption{$\mu = 5, \text{sla objective} = 100ms$}
    \end{subfigure}
    \begin{subfigure}{0.24\linewidth}
        \centering
        \includegraphics[width=\linewidth]{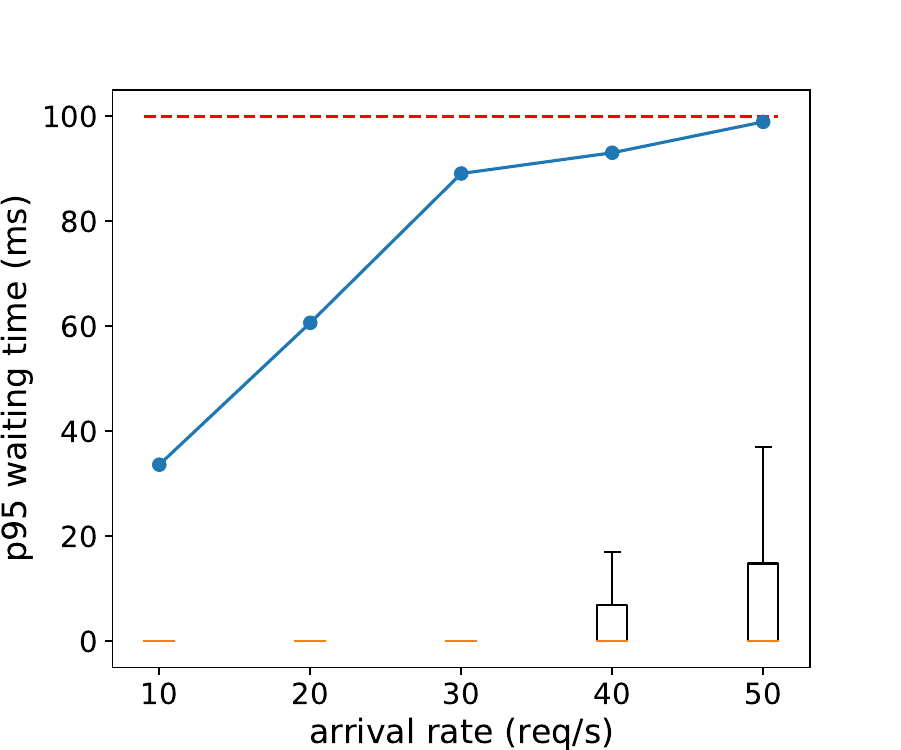}
        \subcaption{$\mu = 10, \text{sla objective} = 100ms$}
    \end{subfigure}
    \begin{subfigure}{0.24\linewidth}
        \centering
        \includegraphics[width=\linewidth]{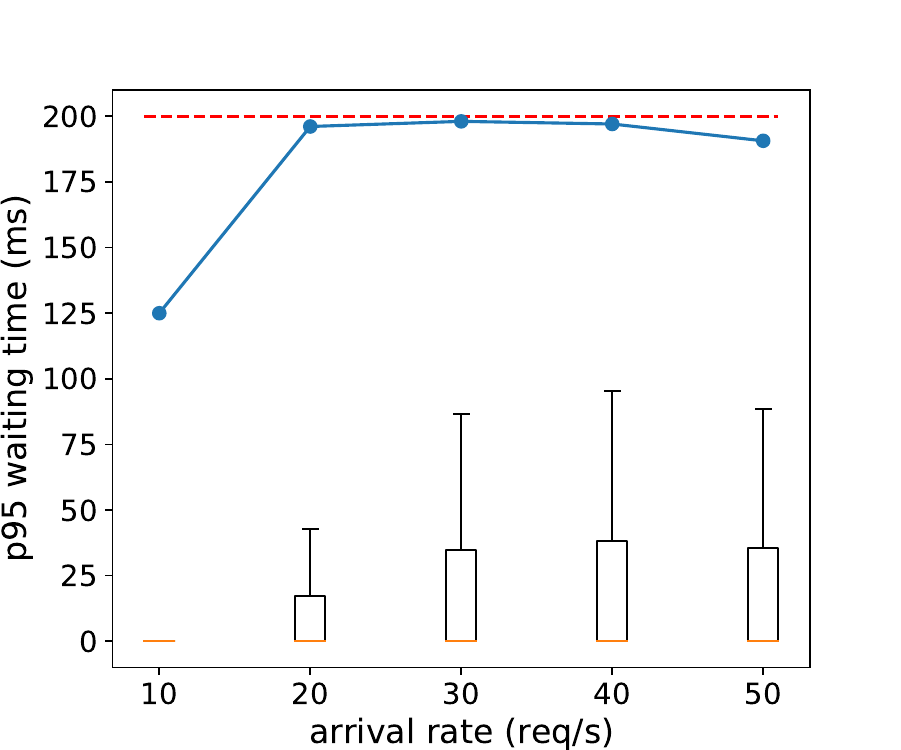}
        \subcaption{$\mu = 5, \text{sla objective} = 200ms$}
    \end{subfigure}
    \begin{subfigure}{0.24\linewidth}
        \centering
        \includegraphics[width=\linewidth]{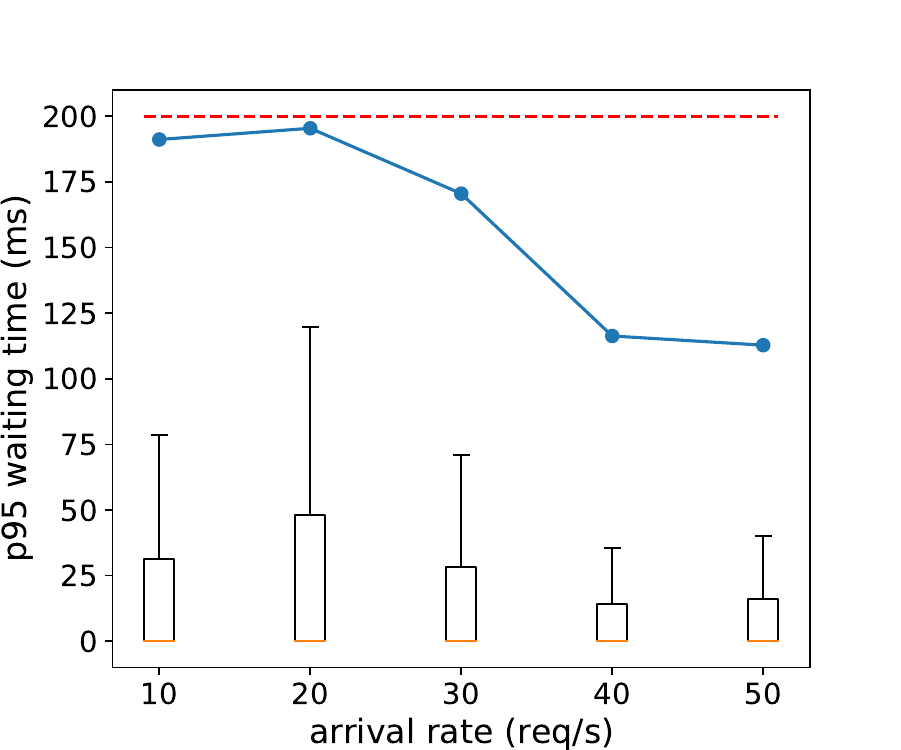}
        \subcaption{$\mu = 10, \text{sla objective} = 200 ms$}
    \end{subfigure}
    \vspace{-3 mm}
    \caption{Measurement of P95 waiting time for functions with homogeneous containers}
    \label{fig:validation-homogeneous}
\end{figure*}

Figure~\ref{fig:validation-homogeneous} shows the required $95^\text{th}$ percentile waiting time
with the red dashed lines, and our empirically measured waiting times when using LaSS with the
$95^\text{th}$ waiting time in blue, along with a box and whiskers plot showing the waiting time
range. As can be seen the empirically observed P95 waiting time are below or close to the SLO
deadline, which shows that our queueing models are able to provision adequate container capacity $c$
for different arrival and service rates while meeting the SLO deadline.

\subsubsection{Heterogeneous Containers}

To validate the queueing model for heterogeneous containers; We run the SqueezeNet function for 10 minutes
under static load using LaSS with no resource constraints to provision just
enough homogeneous containers. Then we randomly select a certain proportion (25, 50, 75, and 100\%)
of all provisioned containers and deflate each selected container randomly. This way, the
function will be under-provisioned with heterogeneous containers and LaSS will react by adding more
containers using the queuing model discussed in \S \ref{sec:heterogeneous_model}. We empirically
measured the waiting time for 20 minutes after the manual deflation.

\begin{figure*}[htb]
    \centering
    \begin{minipage}[t]{.32\textwidth} %
        \centering
        \includegraphics[width=\linewidth]{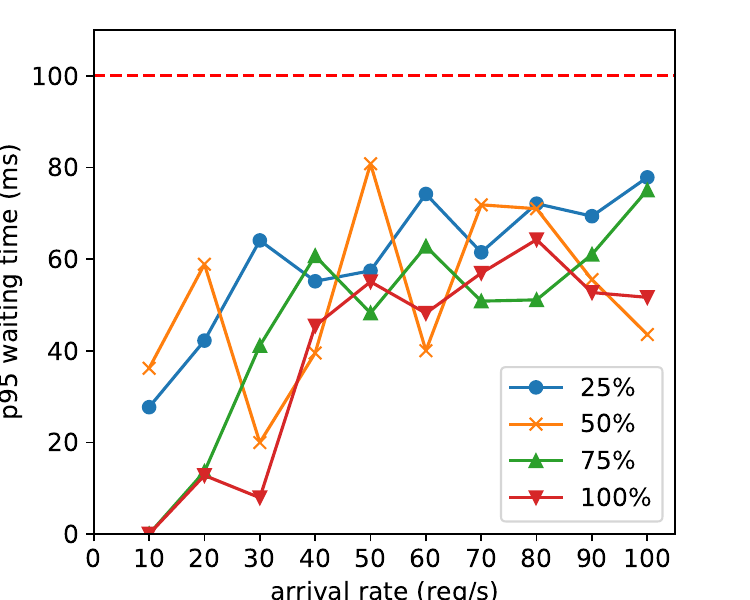}
        \caption{P95 waiting time for functions with heterogeneous containers under different levels of heterogeneity.}
        \label{fig:validation-heterogeneous}
    \end{minipage}\hfill
    \begin{minipage}[t]{.32\textwidth}
        \centering
        \includegraphics[width=\linewidth]{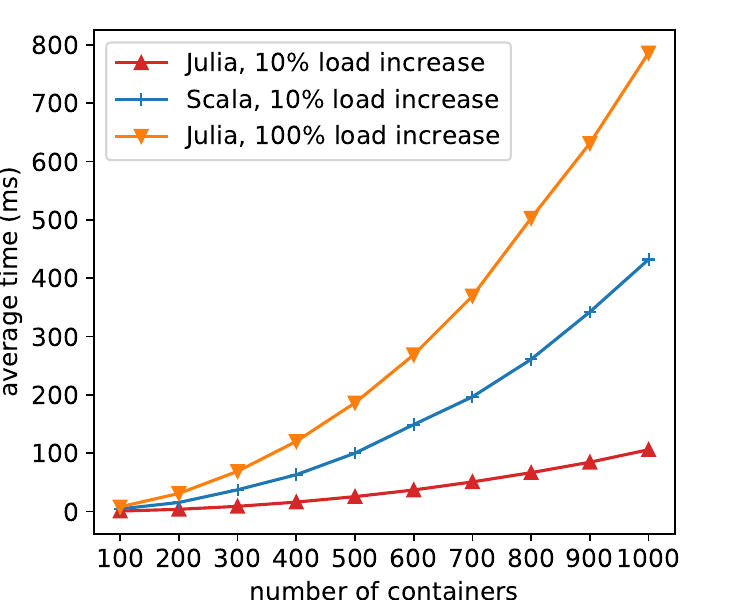}
        \caption{Average computation time of the allocation algorithm for one serverless function with heterogeneous containers.}
        \label{fig:model-scalability}
    \end{minipage}\hfill
    \begin{minipage}[t]{0.32\textwidth}
        \centering
        \includegraphics[width=\linewidth]{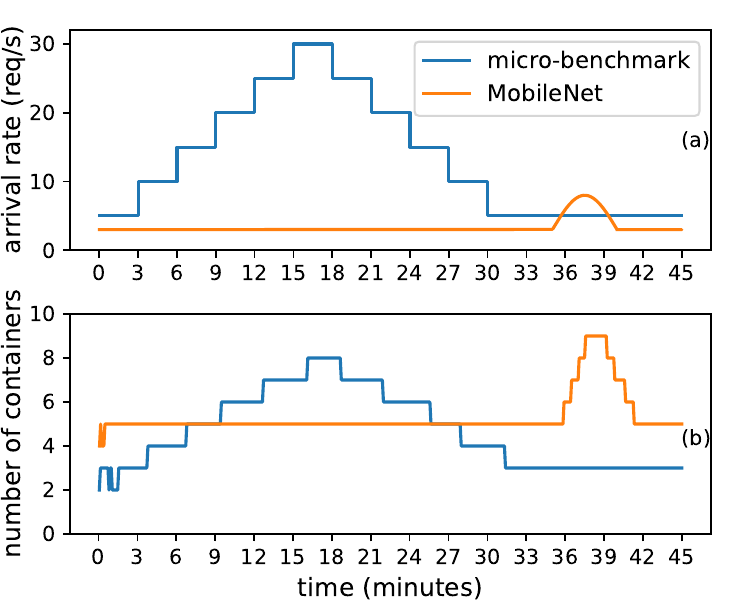}
        \vspace{-5 mm}
        \caption{Two functions with dynamic workloads (a) and the number of containers provisioned by the LaSS over time (b).}
        \label{fig:autoscaling}
    \end{minipage}
\end{figure*}

The results are shown in Figure \ref{fig:validation-heterogeneous}. The x-axis represents the
invocation rates we tested (from 10 to 100 in steps of 10). The different colors of lines represent
the proportion of containers deflated among all provisioned containers. The SLO deadline of 100 ms
waiting time is shown with the red dashed line. As can be seen in all cases LaSS was able to
provision adequate containers to maintain the $95^\text{th}$ waiting time well below the SLO
deadline, which indicates our queueing model for heterogeneous containers is effective.

\subsection{Model Scalability}
\label{subsec:model_scalability}

We now focus on the scalability of allocation algorithm discussed
in \S \ref{subsec:allocation_algorithm}. As the number of applications running at the edge increase,
the number of containers LaSS  manages also increases. Hence, it is important for LaSS to be
able to scale to hundreds or even thousands of containers with sub-second system response time to
handle load bursts. We
implemented the container allocation algorithms in both Scala---since our system is based on OpenWhisk---and Julia, a popular language for scientific computing that accelerates the computations. Figure~\ref{fig:model-scalability} shows for one function, when experiencing a
spike of 10\% increase in request rate, how the number of allocated containers affects the time LaSS
takes to react to the spike. The Figure shows that the Julia implementation of our allocation
algorithm provides much higher scalability, being able to react to spikes within less than 100 ms
even with a 1000 running containers.

Since the burst size can affect the computation time of LaSS, we run an experiment where the
workload doubles instead of gradually increasing. The orange line in
Figure~\ref{fig:model-scalability} shows the time required by LaSS to decide resource allocation
using the Julia implementation (the Scala implementation was not able to compute the results in some
cases due to its precision limitations). We note that while there is a significant increase in the
number of requests, LaSS is still able to scale-up the resources to meet the target SLO in
sub-second time. Since the computation of the allocation algorithms is a relatively separate module,
we were able to integrate our prototype with the Julia module. Note that the the allocation
algorithms can be computed in parallel for different functions. This is an important feature of our
system as it enables LaSS to handle many functions with thousands of containers and react to load
dynamics within a second.

\subsection{Model-driven Auto Scaling}
\label{sec:autoscaling}

Our next experiment shows that the ability of Lass to react to time varying workloads by allocating
an appropriate number of containers to handle an increase or decrease in workload. Our experiment
uses two of our six functions: the micro-benchmark function and the MobileNet v2 deep learning
function. In the first half of the experiment we keep the arrival rate of the MobileNet v2 function
static while the arrival rate of the micro-benchmark function is increased from 5 req/s to 30 req/s
in increments of 5 then decreased back to 5 req/s. In the second half of the experiment, the arrival
rate of the micro-benchmark function is kept static while the arrival rate of the MobileNet v2
function is gradually increased from 3 req/s to 8 req/s then decreased back to 3 req/s. There is no
resource pressure throughout this experiment.

The resulting workload is shown in the upper part of Figure \ref{fig:autoscaling}. The number of
containers allocated to each function is shown in the lower part of Figure \ref{fig:autoscaling}. As
can be seen, in the absence of resource pressure, LaSS can use our queueing models to accurately
estimate the capacity needed and adjust the number of allocated containers in responding to both
sudden and gradual workload increase/decrease. This experiment demonstrates that in the absence of
resource pressure, our LaSS system can quickly react to workload change and auto-scale the container
capacity to ensure low response times.

\begin{figure*}[htb]
    \vspace{-0.5cm}
    \centering
    \begin{subfigure}{0.45\textwidth}
        \centering
        \includegraphics[width=.8\linewidth]{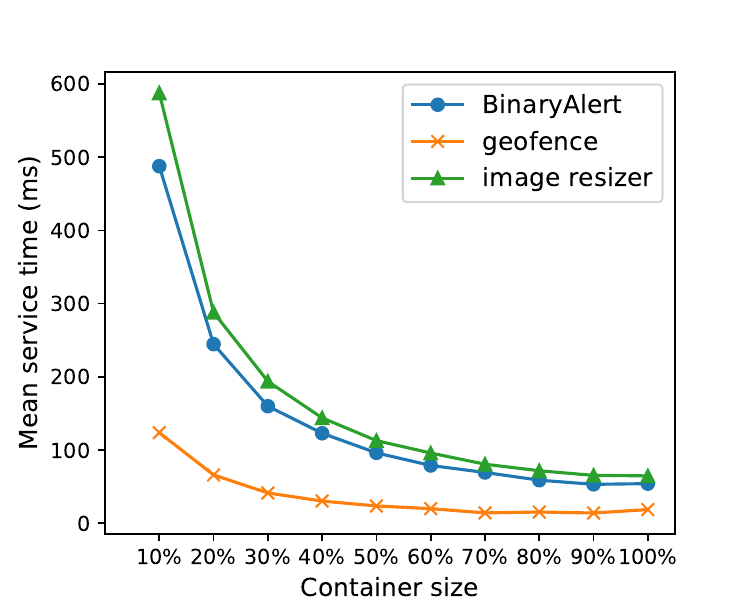}
        \subcaption{Service time of the three non-DNN serverless workloads under different deflation ratios, 100\% container size mean 1 vCPU is assigned to the container.}
        \label{subfig:deflation-non-DNN}
    \end{subfigure}
    \hspace{0.5cm}
    \begin{subfigure}{0.45\textwidth}
        \centering
        \includegraphics[width=.8\linewidth]{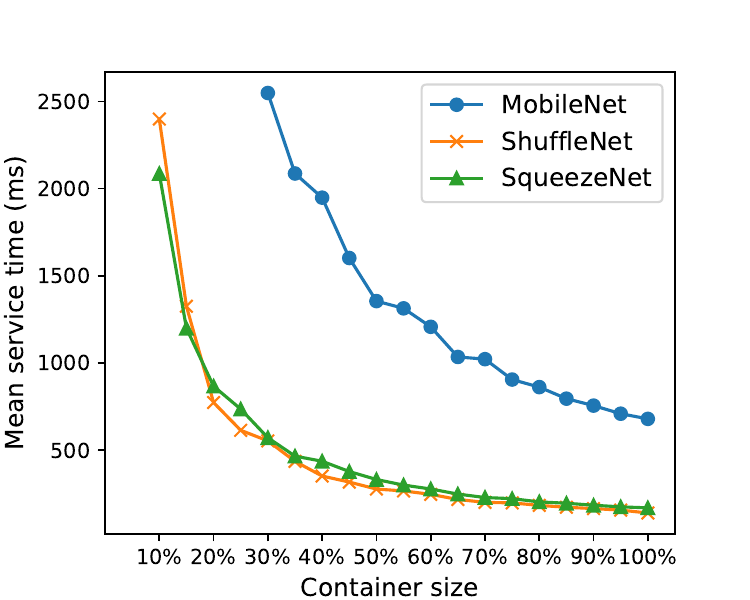}
        \subcaption{Service time of the three deep neural networks under different deflation ratios, 100\% container size means 2 vCPUs are assigned to the container.}
        \label{subfig:deflation-DNN}
    \end{subfigure}
    \vspace{- 3 mm}
    \caption{Effects of deflation on function service time.}
    \label{fig:deflation}
\end{figure*}

\subsection{Efficacy of deflation}

Our next experiment demonstrates the efficacy of container deflation as a viable approach for
resource reclamation. As explained in section \ref{sec:fairness}, normally an application running
inside a container does not use all of the CPU and memory allocated to it. Hence, reclaiming this
spare capacity has negligible or small performance impact. Larger amounts of deflation can however
introduce a proportionate performance impact.

To measure the impact of CPU deflation on performance, we run all six serverless functions inside
containers and progressively deflate the CPU allocation and measure the mean service time.
Figure~\ref{fig:deflation} shows the impact of varying degrees of CPU deflation on the six
functions. As can be seen, for 5 of the functions tested, deflating the CPU by 30\% only yields a
small penalty on service time. As the degree of deflation further increases up to about 70\%, there
is a linear increase in service time.

The MobileNet function shows a slightly different pattern because it is resource heavier compared to
other functions: even if the container is assigned with 2 vCPUs there is little headroom and the
function will run at close to 100\% CPU utilization inside container. This is almost the worst case
for deflation. Still we can see that small amounts of deflation -- around 30\% -- yields roughly a
similar increase in inference time. Even high amounts of deflation -- up to 50\% -- does not come
with any abnormal behavior and sees a corresponding performance decrease.

\subsection{Efficacy of Resource Reclamation}
We now focus on how different resource reclamation policy affects the system utilization. We run two
functions, BinaryAlert malware detection and MobileNet, in the same environment with very high
resource pressure and CPU overload. We assume that both functions are given equal weights $w_1 = w_2
= 1$.

\begin{figure*}[htb]
    \vspace{-0.5cm}
    \centering
    \begin{subfigure}{0.33\textwidth}
        \includegraphics[width=\textwidth]{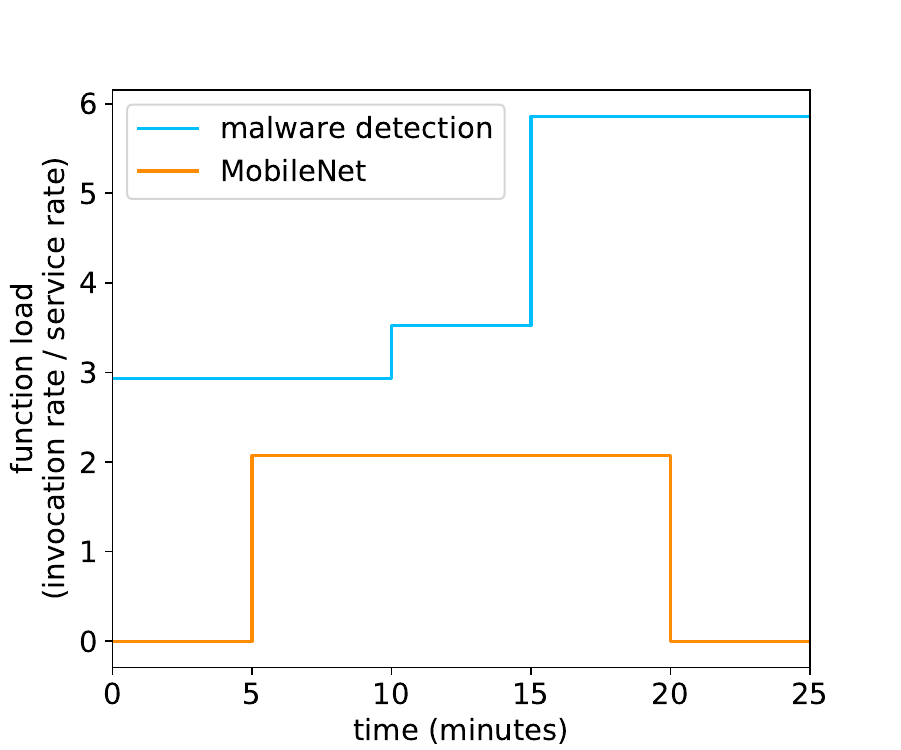}
        \subcaption{Workload seen by MobileNet and malware detection serverless functions.}
        \label{subfig:two-function-trace-overload}
    \end{subfigure}
    \begin{subfigure}{0.33\textwidth}
        \includegraphics[width=\linewidth]{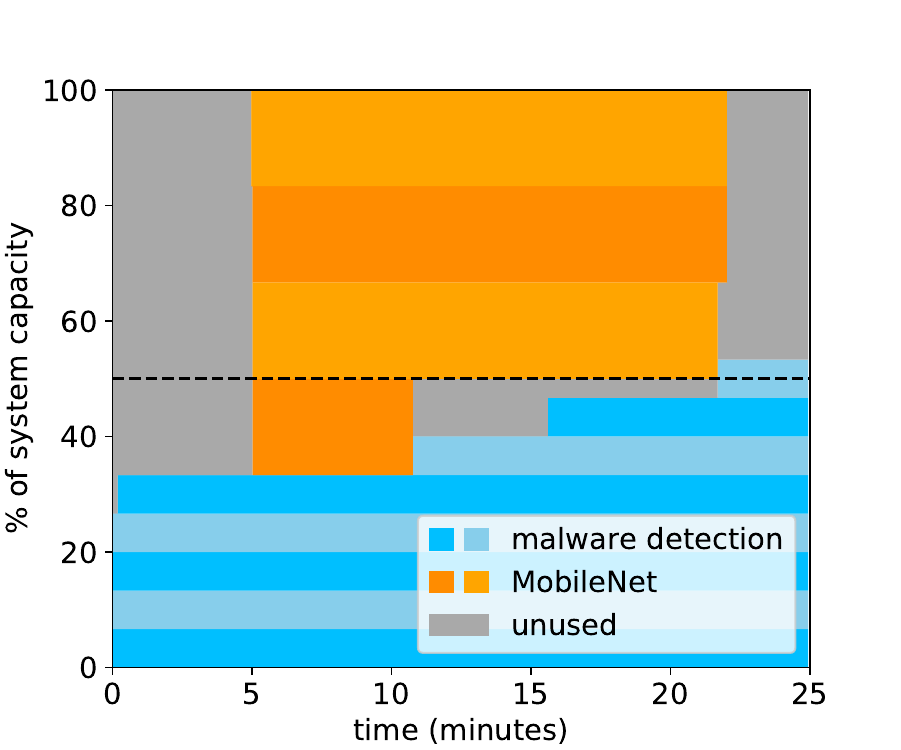}
        \subcaption{Resource allocation to each function using the termination policy.}
        \label{subfig:two-function-allocation-no-deflation}
    \end{subfigure}
    \begin{subfigure}{0.33\textwidth}
        \includegraphics[width=\linewidth]{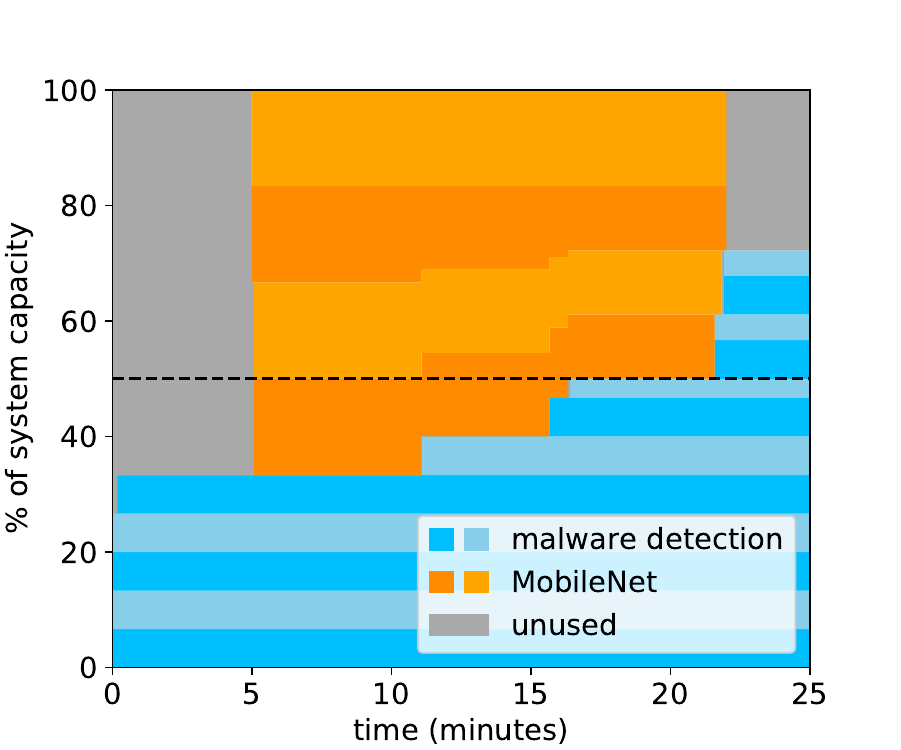}
        \subcaption{Resource allocation to each function using the deflation policy.}
        \label{subfig:two-function-allocation-deflation}
    \end{subfigure}
    \vspace{-3 mm}
    \caption{Effects of different resource reclamation policies under overload, with two functions and synthetic workloads.}
\end{figure*}

Fig \ref{subfig:two-function-trace-overload} shows the workload seen by the malware detection and
deep learning serverless functions. Initially, only the malware detection function is serving
requests and there is no overload. At $t = 5$ minutes, the MobileNet function starts receiving
requests. At this point the malware detection function needs less than its fair share (as its load
hasn't changed) while the MobileNet function needs more than its fair share. At $t =10$ minutes, the
arrival rate of the malware detection function increases which causes the function needing 1 more
container but still less than its fair share. The increasing workload seen by both functions causes
an overload at $t = 10$ minutes. The load of the malware detection function further increases at $t
= 15$ minutes, at which point both functions need more than their fair share. At $t = 20$ minutes
the load of the MobileNet function ceases, which makes all resources in the system available to the
malware detection function.

Fig \ref{subfig:two-function-allocation-no-deflation} shows the behavior using the termination
policy. The different shades of orange and blue represent individual containers and capacity
allocation of each. Under the termination policy, the aggregate capacity allocated to MobileNet is
reduced to 50\% at $t = 10$ by terminating one of the containers and the capacity is reassigned to
the other function, even though the terminated container is much larger in size than the container
to create. At $t = 15$, note that even though the workload of the CPU-intensive function continues
to rise, its share is capped to 50\% which is its fair share under overload. However, since
container deflation is disabled, there is a small fragment of capacity left unused because a
standard sized container of the malware detection cannot fit.

Fig \ref{subfig:two-function-allocation-deflation} shows the system behavior under deflation. In
this case the capacity of the MobileNet containers also needs to be reduced, but this is done by
deflating orange containers. For example, at $t = 10$, the number of containers allocated to the
MobileNet function stayed deflated but three of them were proportionally deflated (depicted as the
width of the of the orange bars becoming narrower) to reclaim just enough capacity to create one new
container for the malware detection function, while still allowing the MobileNet function to use
more than its fair share. Also at $t = 15$ the malware detection function was able to use all of its
fair share allocation using a deflated container. Compared to the termination policy, it can be
observed that there is no unused capacity under the deflation policy, which leads to better resource
utilization. We measure the system utilization improved from 78.2\% when using the termination
policy to 83.2\% (an increase of 6.4\%). Also note that at any time point both functions get at
least the same amount of resources under the deflation policy compared to the termination policy,
which means they always get equal or better performance under the deflation policy compared to the
termination policy.
In both cases, the burst seen by MobileNet ceases at $t=20$ minutes. Since the system is no longer
under resource pressure, our models allocate additional capacity to the malware detection function
shown in blue, allowing it to exceed its fair share in the absence of resource pressure.

We also run the same experiment with off-the-shelf Apache OpenWhisk. However, OpenWhisk failed to
finish the experiment. Soon after the ML workload starts, all invokers become unresponsive. Further
inspection on system logs reveals that the default scheduling algorithm implemented in OpenWhisk
(called the sharding pool load balancer) has caused a cascading failure. By default OpenWhisk tries
to schedule different functions onto different invoker nodes to provide some performance isolation
and maximize chances of container reuse. OpenWhisk also schedules functions solely on memory
requirements while ignoring CPU requirements. Therefore, in this case, one of the invokers is
over-packed with MobileNet containers shortly after the ML workload starts, leaving the invoker
unresponsive. Then again the controller will try to schedule all the ML workload on another invoker,
eventually causing all the invokers to fail. In contrast, LaSS ensures the system can survive
overload by fair share resource allocation and resource reclamation.

\begin{figure*}[htb]
    \centering
    \begin{subfigure}{0.33\textwidth}
        \includegraphics[width=\textwidth]{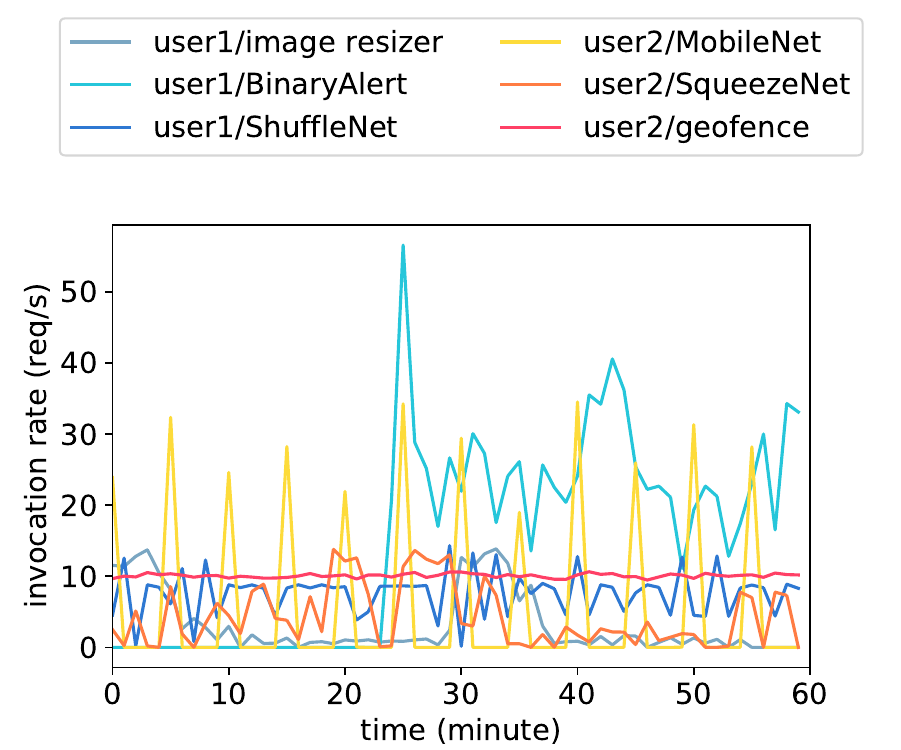}
        \subcaption{Traces for the function invocation rate for the six workloads.}
        \label{subfig:multi_trace}
    \end{subfigure}
    \begin{subfigure}{0.33\textwidth}
        \includegraphics[width=\linewidth]{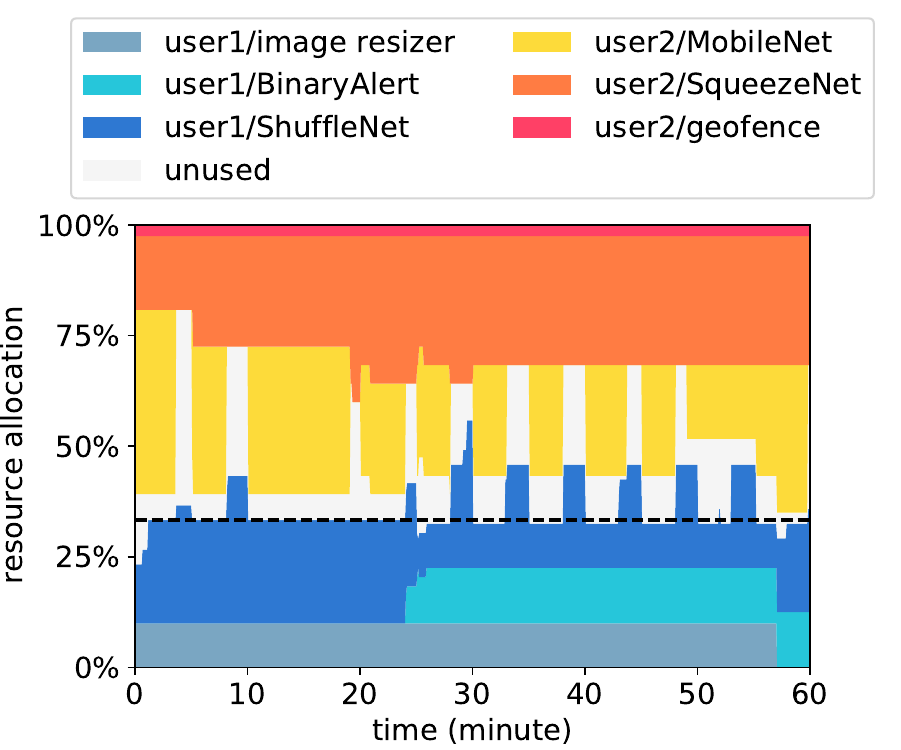}
        \subcaption{Resource allocation to each function using the termination policy.}
        \label{subfig:mult-function-no-deflation}
    \end{subfigure}
    \begin{subfigure}{0.33\textwidth}
        \includegraphics[width=\linewidth]{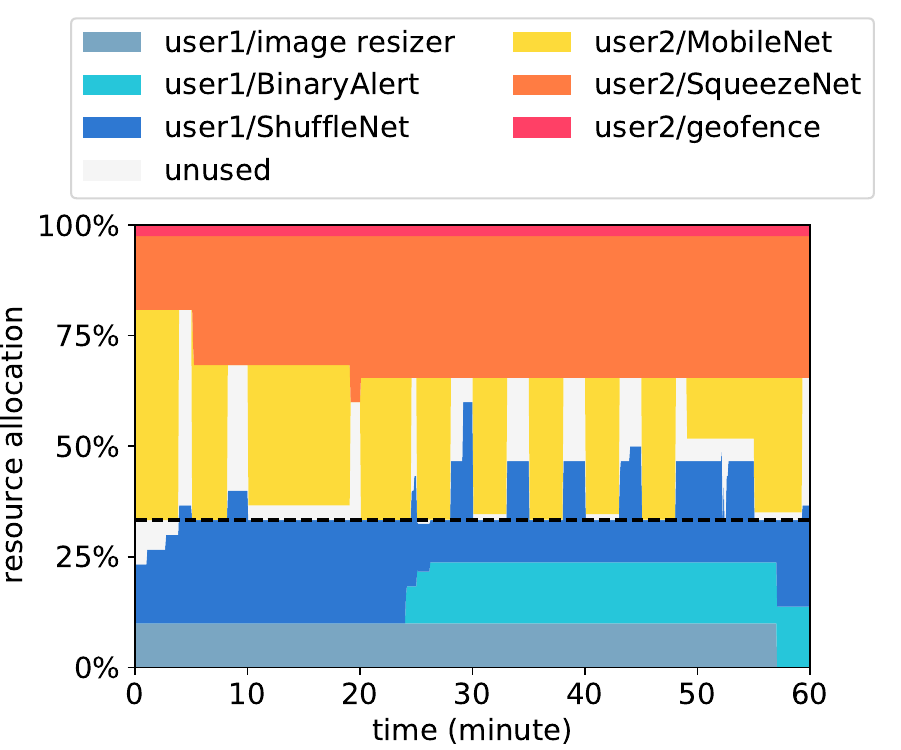}
        \subcaption{Resource allocation to each function using the deflation policy.}
        \label{subfig:mult-function-with-deflation}
    \end{subfigure}
    \vspace{- 3 mm}
    \caption{Effects of different resource reclamation policies under overload, with six functions and workloads sampled from Azure Public Dataset.}
    \vspace{-10pt}
\end{figure*}

\subsection{Function Placement for Azure-like workloads} \label{sec:AzureE}
Finally we want to study how different resource Reclamation policy works with multiple functions. In
this experiment we have all six functions running concurrently, and the entire cluster highly
utilized. We created two users, each running three functions. We set the weight of user 2 to be
twice the weight of user 1. Therefore, when there is resource contention, ideally functions of user
1 can use around 33\% of the total system resources while functions of user 2 can use around 66\% of
the total system resources.

For the workloads in this experiment we use the the Azure Functions Trace 2019
\cite{shahrad2020serverless} from the Azure Public Dataset. The trace is collected in July 2019 and
contains part of the production workload in Microsoft's Azure Functions offering. There are 14 csv
files in the dataset that contain the invocation count of each function recorded over a 24-hour
period, aggregated per minute. For each of the six functions, we sampled a workload for one hour
duration (11:00 am to 12:00 pm in the original dataset) from the Azure Functions Trace.
Figure~\ref{subfig:multi_trace} shows a trace of function invocation calls for all six functions.
It is worth noting here that the MobileNet workload follows a highly sporadic pattern.

Figure~\ref{subfig:mult-function-no-deflation} shows the resource allocation of each function over
time when only the termination policy is applied. The black dashed line represents the ideal
fair-share resource allocation for both users. We can see that when the MobileNet function is not
getting requests, functions of user 1 can user more than their fair-share resource allocation
because the other two functions of user 2 do not need all the fair-share resource allocation of user
2. However, when the MobileNet function starts receiving requests, LaSS quickly terminate some
containers of the ShuffleNet function of user 1 in order to reclaim the resources.

Figure~\ref{subfig:mult-function-with-deflation} shows the resource allocation of each function over
time when the deflation policy is used for resource reclamation. We can see that 1) Under the
deflation policy there is less unused resources (represented by the grey area in the graph) during
overload. We measure the system utilization improved from 87.7\% to 93\% (an increase of 6.1\%).
This is because with deflation there is less resource fragmentation. 2) In Figure
\ref{subfig:mult-function-no-deflation} there are more transient change in allocated capacity
(e.g., around $t = 20$ and $t = 25$ for the SqueezeNet function) compared to Figure
\ref{subfig:mult-function-with-deflation}. This is because under the deflation policy there are much
less container creation/termination operations. From a user perspective this means the service can
seem more stable under the deflation policy since there are less cold starts (due to container
creation) and fewer requests that need to be rerun (due to container termination). 3) Also note that
under the deflation policy all the functions get at least the same amount of resources as under the
termination policy, and different reclamation policies have almost no affect on functions whose
required resource allocation doesn't exceed their fair share.
\vspace{-1 mm}

\section{Related work}

\noindent\textbf{Serverless runtimes and platforms.}  While Docker is currently the most popular runtime for
serverless functions, other runtimes have emerged such as microVMs~\cite{agache20firecracker}, and
unikernels~\cite{tan20unikernel}. Running on top of these runtimes, many serverless platforms have been designed and built by the research community~\cite{akkus2018sand, koller2017will,
chard2020funcx}, including building real-time serverless platforms that provides an invocation rate
guarantee, a service-level objective (SLO) specified by the application, that is delivered by the
platform~\cite{nguyen2019real}. Other platforms focus on the problem of state and data management in
the serverless paradigm~\cite{klimovic2018pocket,zhang2019narrowing}, which is one of the major
weaknesses of the serverless paradigm~\cite{hellerstein2018serverless,wang2018peeking}. Due to the short-lived nature of
serverless functions, many of the assumptions on data locality that modern CPU architectures make
resulting in larger switching and hardware caching overheads which can result in up to 20x slowdowns
compared to native execution~\cite{shahrad2019architectural}. However, many case-studies have shown
that serverless can save hosting cost by up to 95\% for web-services~\cite{10.1145}, and provide
order of magnitude performance speed-ups~\cite{3307339}.

\noindent\textbf{Scheduling and management of serverless functions.} \texttt{gg} is a framework that
enables users to deploy, execute, and manage applications using serverless functions deployed on
thousands of parallel threads to achieve near-interactive completion times~\cite{234886}. MPSC is a
framework for scheduling serverless functions across different cloud providers based on the
performance of each provider in a given time~\cite{aske2018supporting}.
FnSched~\cite{suresh2019fnsched} is a scheduler implemented on top of OpenWhisk to regulate the
resource usage of co-located functions on each invoker in the system. Recent studies have shown the
extremely high variability in invocation rates for serverless functions with variations of 8 order
of magnitude in the rate~\cite{shahrad2020serverless}. This problem is related to the bursty
workload problem in traditional cloud environments~\cite{tai2011ara,issawi2015efficient}. Using
queueing theory-based approaches to model and manage the performance of distributed systems have
been extensively studied \cite{harchol2013performance,urgaonkar2005analytical,
gandhi2014adaptive,wierman2009power,khazaei2011performance}. Several papers have considered
improving system performance by employing different scheduling policies, e.g., the Shortest
Remaining Processing Time (SRPT) scheduling policy instead of FCFS due to its optimal properties
\cite{bansal2001analysis, harchol2003size}.

\noindent \textbf{Serverless real-time applications.} While relatively new, serverless computing is
today being used for many real-time applications. A popular class of applications that use
serverless computing are applications that include machine learning inference~\cite{zhang2019mark}.
Another popular class of applications are those involving real-time video streaming
processing~\cite{zhang2019video,ao2018sprocket}. Sprocket~\cite{ao2018sprocket} is a serverless
based video processing framework supporting both batch and streaming video processing.
Lavea~\cite{yi2017lavea} is a serverless-based edge video analytic platform that is capable of
providing between 1.3x to 4x  speedups compared to running the analytics locally.

\section{Conclusion}
We presented LaSS, a serverless platform designed for latency sensitive computations
at the edge. LaSS uses queueing theory based models to determine the container capacity needed by a
latency sensitive function to meet its deadlines. LaSS also ensures fair share resource allocation
when the system is overloaded and can reclaim resources from over-provisioned functions using
different reclamation policies. We implemented a prototype system on top of Apache OpenWhisk.
Experimental results indicate that our models can make accurate predictions of required capacity for
latency sensitive workloads, and when the system is overloaded each function will get its guaranteed
fair share with our resource allocation algorithm. We also show that deflation can lead to better
resource efficiency compared to termination when resource reclamation is needed.

There are a number of directions in which to extend this work. For example, we have only considered
Poisson arrival and service processes. We can generalize our models to other inter-arrival/service
time distributions. Another direction for further investigation is to take composition of serverless
functions into consideration when making scheduling and resource allocation decisions
\cite{baldini2017serverless, shahrad2020serverless}, so that functions that belong to the same
application can be scheduled in a coordinated manner.

{\noindent \bf Acknowledgements.}  We thank the anonymous reviewers and our shepherd for their helpful comments. This research was funded in part by NSF grants 1908536, 1836752, 1763834, US Army contract W911NF-17-2-0196, Chalmers ICT-AoA, and Amazon AWS cloud credits.

\bibliographystyle{ACM-Reference-Format}
\balance
\bibliography{main}
\end{document}